# Investigation of plasmonic evolution of atomically size-selected Au clusters by electron energy loss spectrum—— from solid state to molecular scale

Siqi Lu[1†], Lin Xie[2†], Kang Lai[3†], Runkun Chen[4,5†], Lu Cao[1], Kuojuei Hu[1], Xuefeng Wang[6], Jinsen Han[3], Xiangang Wan[1], Jiaqing He[2*], Jiayu Dai[3*], Jianing Chen[4,5,7*], Qing Dai[8], Zhenlin Wang[1], Guanghou Wang[1], Fengqi Song[1*]

1. National Laboratory of Solid State Microstructures, Collaborative Innovation Center of Advanced Microstructures, and School of Physics, Nanjing University, Nanjing 210093, China

2. Department of Physics, Southern University of Science and Technology, Shenzhen 518055, China

3. Department of Physics, National University of Defense Technology, Changsha 410073, People's Republic of China

4. Institute of Physics, Chinese Academy of Sciences and Beijing National Laboratory for Condensed Matter Physics, 100190 Beijing, China

5. School of Physical Sciences, University of Chinese Academy of Sciences, Beijing 100049, China

6. School of Electronic Science and Engineering and Collaborative Innovation Center of Advanced Microstructures, Nanjing University, Nanjing 210093, China

7. Songshan Lake Materials Laboratory Dongguan, 523808 Guangdong, China

8. Division of Nanophotonics, CAS Center for Excellence in Nanoscience, National Center for Nanoscience and Technology, Beijing 100190, China

**Abstract**

Versatile quantum modes emerge for plasmon describing the collective oscillations of free electrons in metallic nanoparticles when the particle sizes are greatly reduced. Rather than traditional nanoscale study, the understanding of quantum plasmon desires extremal atomic control of the nanoparticles, calling for size-dependent plasmon measurement over a series of nanoparticles with atomically adjustable atom number (*N*) over several orders of magnitude. Here we report the *N*-dependent plasmonic evolution of atomically size-selected gold particles with *N* = 100-70000 using electron energy loss (EEL) spectroscopy in a scanning transmission electron microscope. The EEL mapping assigns a feature at ~2.7 eV as the bulk plasmon and another at ~2.4 eV as surface plasmon, which evolution reveals three regimes. When *N* decreases from 70000 to 887, the bulk plasmon stays unchanged while the surface plasmon exhibits a slight red shift from 2.4 to 2.3 eV. It can be understood by the dominance of classical plasmon physics and electron-boundary scattering induced retardation. When *N* further decreases from 887 to 300, the bulk plasmon disappears totally and the surface plasmon shows a steady blueshift, which indicates that the quantum confinement emerges and modifies the intraband transition. When *N* ~ 100 - 300, the plasmon is split to three fine features, which is attributed to superimposed single-electron transitions between the quantized molecular-like energy level by the time-dependent density functional theory calculations. The surface plasmon's excitation ratio has a scaling law with an exponential dependence on *N* ($\sim N^{0.669}$), essentially the square of the radius. A unified evolution picture from the classical to quantum, molecular plasmon is thus demonstrated.

**Introduction**

As an elementary type of collective excitation, plasmon has been found to dominate the optical properties of metals ever since the first experiments were conducted in this area[1], and further interest then arose following the emergence of nanotechnology[2], and in connection with explanations of the Lycurgus Cup. Intense efforts have led to the discovery of some striking behavior, including the existence of hot spots with field enhancement[3,4], coupling-induced optical shifts[5], and geometrically influenced plasmon absorption[6], as well as potential applications such as biological labeling[7,8], infrared waveguides[9,10], cavity[11,12] and quantum-dot[13] displays. The size-dependence of nanoparticle plasmons is of key interest in studies of this type[14-17], given that it not only provides reliable nanoparticles with a standard optical response for subsequent assembly and optical operation[18,19], but it is also the only means of reaching a unified understanding in the broad physics that spans from solid state plasmon in large particles[20], to mesoscale and atomic/molecular scale plasmon in particles with countable atoms[21-23]. Large nanoparticles are believed to have similar electronic structures as bulk metal, and the collective resonance of free electrons persist despite the retardation and relaxation that occurs due to particle size[24,25]. The reduction in the number of atoms makes the classical plasmon model give way to the quantum corrected model and a range of concepts, including the quantum plasmon[14,26-28] and electron spill-out effects[29-32] have emerged during these size-dependent studies. The smallest particles with countable atoms will show quantized molecule-like behaviors[33-35], where the electrons may even be totally localized and plasmonic excitation seems precluded totally[36]. Even more controversy exists on such interesting questions as the division between the nanoparticle and molecules[15,37], and the physics of mesoscopic and microscopic plasmonic evolution[38-40]. A unified understanding covering three macro / meso / micro scales is required; however, there is still a lack of samples with sufficient atomic precision, covering the whole regime from macro to molecular scale, given that in practice this requires atomically precise samples covering several orders of atomic number ($N$).

To try to shed light on these issues, we prepared mass-selected gold clusters $Au_{100-70000}$ and measured their plasmonic evolution by electron energy loss spectroscopy (EELS) in a scanning transmission electron microscope (STEM). Two peaks were identified at the center and edge of the clusters, allowing us to study the physics of the evolution of their atomic-number dependence. Three regimes with distinct plasmonic physics were observed. $Au_{887}$ was found to be at the boundary between the classical plasmon of nanoparticles and the quantum confinement corrected plasmon (QCC plasmon). The plasmon related to quantized molecular energy levels (molecular plasmon) arises below $Au_{300}$ and found to be superimposed coherent single-electron transitions.

**Acquisition of high-quality plasmonic signals of individual clusters by STEM-EELS**

Atomically-precise gold clusters were produced using a newly built high-vacuum magnetron sputtering cluster beam source equipped with a lateral time-of-flight mass selector, which guarantees a mass selection resolution of M / ΔM ≈ 50 in a very large mass range[41,42]. All the clusters were deposited onto an ultra-thin carbon film (~ 3 nm) on TEM grids in a soft landing condition, with a beam energy of less than 0.5 eV / atom[43]. Spectroscopic analysis of the deposited gold clusters was performed with a FEI Titan transmission electron microscope operated at 60 kV in STEM mode with an imaging spatial resolution of 0.30 nm and an energy dispersion of 0.01 eV per pixel. The full-width at half maximum (FWHM) of the EELS zero-loss peak is ~0.12 - 0.13 eV. **Figure 1(a)** shows the micrographs of the sample clusters for gold atoms N from 100 to 70000. It should be noted that their diameters increase from ~ 1.5 nm to ~ 15 nm even though their atomic numbers increase by a factor of 700. Particularly, the diameters vary rather inapparently at all when *N* increases from 300 to 600, while the plasmon peak changes considerably as shown below. This indicates the necessity of this atomically precise N-dependent plasmonic study.

An incident high-energy electron will excite the plasmonic resonance of a metallic nanoparticle resulting in a loss of its energy[44,45]. Fig. 1(b) shows a typical electron energy loss spectrum obtained from a gold nanoparticle in the inset, where a peak can be seen clearly at 2 - 3 eV, although its intensity is much lower than the zero-loss peak. This has been attributed to the plasmonic response of gold nanoparticles[46]. As well as the application of intense binning and careful damage control during the spectrum accumulation, the beam-focusing configuration was also optimized as shown in Fig. 1(c), where the plasmonic peak increased tens of times when the camera length was changed from 30 to 73 mm. However, this cameral length is still very small in consideration of dipole approximation to satisfy the existing surface dipole mode for very small nanoparticles[47]. The experimental results show a satisfactory spatial resolution thanks to the use of an aberration-corrected STEM as shown in Fig. 1(d). The edge and center of an $Au_{70000}$ cluster provide different spectroscopic features, where the green curve obtained from the particle edge is assigned to the surface plasmon peak (SPP) (~ 2.4 eV) and the black curve obtained from the particle center is assigned to the combine of SPP and bulk plasmon peak (BPP) (~ 2.7 eV)[48]. In a smaller particle such as $Au_{600}$, no obvious difference can be seen between the spectra from the center and edge of the particle, and it seems that only SPP remains. This abrupt diminish of bulk plasmon can be explained by the different quantization of the two mode influenced by the cluster surface[49], which we will discuss below. The STEM probe is demonstrated to provide a reliable plasmonic signal of individual ultra-small clusters with a high spatial resolution and a satisfactory yield.

**$N$ dependent evolution of the plasmon peaks**

Using this specific electron probe, we measured all the gold clusters with $N$ from 100 to 70000 and analyzed the evolution of their plasmonic peaks. **Figure 2(a)** shows the SPP evolution of the energy loss spectra collected at the edge of all the gold clusters, where the dashed line indicates the variation in the center of the peak. It is located at around 2.4 eV when $N$ is 70000 with the particle diameter about 15 nm. Its position is similar to those of the SPPs of other large gold nanoparticles[50], indicating that these experimental results may show some connection with the classical understanding of surface plasmon resonance in a quasi-static approximation. When $N$ decreases from 70000, the SPP shows a very slight red shift. It changes very slowly from 2.4 to 2.3 eV over the very large scale of $N$ down to $N \sim 887$. With the further decrease of $N$, the SPP shows a gradual blue shift from 2.3 to 2.7 eV down to $N \sim 300$. After that, new modes arise, resulting in three fine features down to the smallest cluster for $N \sim 100$. The three features appear for 1.9 - 2.1 eV, 2.45 - 2.6 eV, and 2.8 - 2.95 eV respectively. Fig. 2(b) shows the evolution of the central excited plasmon energy loss spectra collected in the centers of all the gold clusters, and both SPP and BPP can be detected in the central excitation. The excited plasmon peaks stays steady when $N$ decreases from 70000 to 1100 and then the part located at about 2.7 eV disappears abruptly for smaller clusters with $N < 887$. The remaining peak exhibits blue shift when $N < 887$. For these small clusters, the characteristics of the remaining peak that appear in the same region and exhibit the same blue shift as SPP, and show no difference between the central and edge excitation (fig. 1(d)), which must therefore originate from SPP. We are thus convinced that the BPP disappears for all clusters with smaller $N$.

As expected, the spectra yield of the clusters decreases with the decrease in cluster size, which was indeed observed during the measurements. We calculate the exciting probability of BPP ($P_{BPP}$) by dividing the area of the BPP by the area of zero-loss peak[45]. As shown in Fig. 2(c), $P_{BPP}$ decreases with the decrease of $N$. $P_{BPP}$ decreases by several orders of magnitude and falls to around zero after $N \sim 887$, thus confirming the disappearance of the BPP mode at $Au_{887}$. By dividing the area of the SPP by the area of the zero-loss peak of the whole cluster boundary area, we can also obtain the exciting probability of the surface plasmon ($P_{SPP}$) by a single electron[45]. We find that $P_{SPP}$ decreases rapidly with decreasing $N$. By showing the data in an *ln-ln* framework in Fig. 2(d), all the data fall scattered around a straight line with a slope of $0.669 \pm 0.043$. This implies a simple power law of the form $P_{SPP} \sim N^{0.669 \pm 0.043}$. Assuming that the volume of an atom in the clusters does not change much in any of the clusters, we know $N$ is proportional to $R^3$ (the radius: $R$), i.e. $P_{SPP} \sim R^{2.007 \pm 0.129}$, where the exciting ratio of the SPP at the edge of the clusters depends on the area of the gold clusters. This confirms the surface origin of the observed SPP. We also measured the full width at half maximum (FWHM) of the SPP, as shown in Fig. 2(g), which shows a large decrease for small clusters.

Fig. 2(e-g) summarize the data obtained in three characteristic regimes separated by the two grey bold vertical dashed lines. In regime 3 ($N$ ~887-70000), the positions of the SPP exhibit a very slight red shift with decreasing $N$, while the BPP remains unchanged and the FWHM remains at a high value of about 0.36 eV. In regime 2 ($N$ ~300-887), the position of the SPP exhibits a steady blueshift with decreasing $N$, while the BPP disappears altogether and the FWHM stays at a high value of about 0.26 eV. In regime 1 ($N$ ~100-300), the SPP shows 3 fine features with a much smaller FWHM, which is close to the FWHM of the EELS zero-loss peak. The physics of the three regimes is discussed below.

**From electron-boundary scattering modified classical plasmon to QCC plasmon at Au$_{887}$**

Over a large range of N (887-70000), the SPP exhibits a slight red shift with decreasing N while the BPP is maintained fairly well, which can be interpreted in terms of the classical plasmon with a modification after considering electronic scattering by the particle surfaces, as discussed below.

According to classical plasmonic physics, the polarizability of the dipole resonance of the localized surface plasmon is determined by the dielectric functions[51,52]. For large nanoparticles without any obvious difference in electronic structure from those of bulk metals, the reduction in the size only introduces more scattering due to the boundaries for free electrons in the metal, besides the Coulomb scattering between electrons[24,25], as shown in **Figure 3(a)**. Hence the dielectric function can be formulated[53] by : $\varepsilon(\omega) = \varepsilon_\infty - \frac{\omega_p^2}{\omega(\omega+i\gamma)} = \varepsilon_\infty - \frac{\omega_p^2}{\omega(\omega+i\gamma_{bulk} + \frac{iAv_F}{R})}$, where $\varepsilon_\infty$ is incorporated in the dielectric function considering background electron screening at high frequency, and the final term in the denominator shows a correction due to boundary scattering. In essence, $\gamma = \gamma_{bulk} + \frac{Av_F}{R}$ represents a correction to the electron scattering probability. $v_f$ is the Fermi velocity of free electrons in gold nanoparticle. A is an empirical constant and it takes a value from 0.1 to 2, which is an empirical constant that takes into account the details of the scattering processes. For example, A = 0 for elastic scattering, A = 0.75 for diffuse scattering, A = 1 for isotropic electron scattering, A > 1 if there are any additional limitations imposed by internal grain boundaries and this additional scattering process simply indicates the surface scattering by electrons (volume of the cluster) with the surface of the cluster[1,54-56]. As shown in Fig. 3(b), we fit the experimental data of the *N*-dependent SPP data using the above formula, which reveals a good agreement when A = 0.4 or 0.5 which is between the value 0.25[57] and 0.7[58] reported, and we confirm the contribution of the boundary scattering influenced by the different structure and surface electrons density of the cluster[1].

The electronic conduction band, valid at macroscopic scales, breaks down with some gaps when the dimensions are small enough. This quantum confinement effect makes the classical Drude model for the dielectric function invalid[27]. In our experiment the classical condition may fail for clusters with smaller *N* (~887) when their electronic structures are strongly affected by quantum confinement, given that there are so few atoms[59]. The resulted failure of the conduction band of the gold nanoparticles naturally leads to the observed disappearance of the BPP[49]. However, the surface electrons are more diffusive, which have a stronger softening effect on the quantization of the surface mode than the bulk mode, resulting in a quantum confinement corrected SPP because fewer electronic or plasmonic transitions are allowed. The permittivity of the Au clusters can be calculated using the Drude model modified with quantum confinement effects[14,60]. The total permittivity $\varepsilon_{Au}$ is the sum of the permittivity of free electron transitions in the quantized conduction band and the frequency-dependent permittivity of interband transitions between the d bands and the higher conduction bands[50], whose contribution to the dielectric function in noble metals is dominated by the outermost d electrons in the atoms concerned[61]. By replacing $\varepsilon_\infty$ with the frequency-dependent $\varepsilon_{inter}$ in the classical dielectric function and considering a quantum-modified conduction band[14], the total particle permittivity can be formulated as: $\varepsilon(\omega) = \varepsilon_{inter} + \omega_p^2 \sum_i \sum_f \frac{s_{if}}{\omega_{if}^2 - \omega^2 - i\omega\gamma}$, where $\omega_{if}$ is the frequency of the transitions in the conduction-band electrons from the initial state i to the final state f; $S_{if}$ is the oscillator strengths of the corresponding transitions, and the scattering frequency $\gamma$ is modified by the effective electron mean free path influenced by the boundary scattering, which depends on the radius R[24,25]. For the simplest case, we consider only the strongest transitions that emanate from states at the Fermi surface and can get an approximate solution in a simplified formula[62]

$$\omega_s^2 = \omega_p^2 \left( \frac{\omega_{qm}^2 - \gamma^2}{\omega_p^2} + \frac{1}{\varepsilon_{inter}^{Re}(\omega_s) + 2\varepsilon_m} \right),$$

which is valid for a defined size-dependent dipole transitions $\omega_{qm}$ at Fermi surface, representing the quantized conduction band which is described by $\omega_{if}$ and $s_{if}$ above[63]. In a simple box model[62,63] the corresponding quantum induced blue-shifted energy $\hbar\omega_{qm} = \hbar\omega_p \frac{R_0}{R}$, $R_0$ can be written in terms of the effective free-electron density parameter $r_s$ as $R_0 = 1.1 a_0 \sqrt{r_s}$, and $a_0$ is the Bohr radius. Good agreement can be seen in Fig. 3(c). The corresponding energy gap at the Fermi level obtained by Kubo[64] ranges from about 5 meV for $Au_{887}$ to 20 meV for $Au_{300}$. This confirms the transition from the classical plasmon to QCC plasmon when (N ~887-300). We note that $Au_{887}$ is very small, which indicates little quantum effect for most gold nanoparticles since they are larger than $Au_{887}$. Finite element calculations based on classical electromagnetism and the dielectric function may thus be used to tackle most nanoplasmonic designs using the present industrial nanofabrication.

**Superimposed transitions between quantized molecule-like electronic structures ($N < 300$)**

With only a few dozens to hundreds of atoms, bulk electronic structure is expected to give way to complex molecular one due to quantum nature and serious debates have been raised as to the electronic response nature of these small clusters[65]. As seen in fig. 2 (e), for $N < \sim 300$, three fine structures arise in 1.9 - 2.1 eV, 2.45 - 2.6 eV, and 2.8 - 2.95 eV respectively, which are out of the traditional understanding of plasmon physics with a single spectroscopic feature[66]. The FWHM of the new features are several times narrower, which indicates these might be from some molecular-like electronic structures. The low values of $N$ means that it is possible to perform real-time time-dependent density functional theory (rt-TDDFT) calculation to understand the plasmonic physics of clusters with a hundred gold atoms from the first-principle level. The details of the calculations can be found in the method section.

We use $Au_{116}$ as a typical example, the atomic structure of which is optimized as shown in **Figure 4(a)**[67]. Its electronic structure is calculated as shown in Fig. 4(b). It is shown that although it exhibits band-like structure around -3 eV, we can also find there are obviously discrete energy levels in the (-2, 4 eV) region with the Fermi level as 0. This clearly shows the molecule-like character of the small gold cluster. We employ the method developed by Wang et al.[68] based on the rt-TDDFT to analyze the origin of the absorption transition. This method has been used for investigating the interplay between plasmon and single-particle excitation in $Ag_{55}$.[31] In practice, after the femtosecond laser pulse is applied to the cluster as shown in Fig. 4(c) (black line), the dipole moments, shown in Fig. 4(c) (red line) will continue to oscillate even at the time of 50 fs, indicating there should be some plasmon modes in the small clusters.

We then calculate the optical absorption spectrum as shown in Fig. 4(d), which may produce spectra contribution in most region of interest of the three features of small gold clusters, as shown in Fig. 4(e). We note that there is some detailed difference between the calculations and experiments. First of all, density functional theory is known to underestimate the energy of excited level, resulting in different positions of the spectra peaks compared with the experiments. Secondly, since we used laser field to excite the electrons, there might be some difference in the STEM experiments[69].

In order to analyze the contribution to spectra peak, the time-dependent transition coefficients $|C_{j,i}(t)|^2$ for the main absorption peaks are checked. When we used the laser field which is weak enough ($0.0257\ V/Å$) to promise the linear excitation of electrons, the transitions could be clearly classified into two different types. One is shown to be rapidly oscillating and another is found to be slowly varying with time evolution, corresponding to plasmon and single-particle transition mode, respectively.[31] We can clearly judge that the optical feature at 1.53 eV is plasmonic absorption while the experimentally-observed features at 2.09, 2.30 and 2.66 eV are single electron energy level excitations. However, things change when we apply a higher electrical field. Upon a laser field of 0.514 V/Å, the transition coefficients of all the experimentally-observed three features will include both single electron excitation and collective electron oscillation modes. Even after the laser driven, the time evolution electron charge density can also find similar dynamics (see supplementary Figure. S2, 3). It means the occurrence of the plasmon oscillation is found at stronger laser field. When the laser filed is stronger, many-electron excitation can happen, and the excited electron may also exhibit collective oscillations. This reveals that the present experimental plasmon is some superimposition of single electron transitions between the quantized molecular energy levels. Given the fact that the STEM-EELS is normally used to probe plasmon rather than single-electron transition and in view of the estimated electrical field of ~$10^9$ V/m around 1 nm of a 60 kV electron beam in vacuum in the present study, we are convinced that the three observed features represent molecular plasmon.

**Conclusions**

Measurements of the *N*-dependent evolution of both the BPP and SPP of size-selected gold clusters Au$_{100-70000}$ were obtained using the STEM-EELS approach. Three regimes were observed, each with distinct physics. In the third regime ($N \sim 887 - 70000$), the SPP exhibits a slight red shift due to gradually apparent electronic scattering by the particle surfaces. In the second regime ($N \sim 300 - 887$), the SPP exhibits a steady blue shift and the BPP disappears altogether due to the quantum confinement effect. In the first regime ($N \sim 100 - 300$), the SPP splits into 3 fine features with very small FWHM, indicating the dominance of the molecular energy levels. A unified set of observations from solid-state classical plasmon physics, QCC plasmon physics, and molecular plasmon are thus demonstrated. This paves the way for new developments in physics and for future applications of nanoplasmonics.

**Materials and Methods**

**Sample preparation**

Gold nanoclusters were produced using a magnetron sputtering gas phase condensation cluster beam source. A time-of-flight mass filter was used to select a cluster of specific atomic numbers, offering a mass resolution of M / ΔM ≈ 50. The mass-selected gold clusters were focused into the deposition chamber under high vacuum conditions ($10^{-5} - 10^{-4}$ Pa), and were deposited onto the ultra-thin carbon film (~ 3 nm) on TEM grids at a soft landing energy of less than 0.5 eV / atom. In this way, Au clusters from 70000 to 100 atoms with the mass resolution M / ΔM ≈ 50 were soft-landed on the substrate.

**STEM EELS collection and data processing**

Spectroscopic analysis of the deposited gold clusters was performed with a FEI Titan transmission electron microscope at 60 kV in STEM mode, with an imaging spatial resolution of ~0.30 nm, an energy dispersion of 0.01 eV per pixel, and an EELS zero-loss peak (ZLP) full-width at half maximum of 0.12-0.13 eV. We found that 60 kV is more suitable for EELS analysis because of the higher excitation probability compared with that of a 300 kV electron energy. Each cluster was mapped in a fully-covered square box evenly divided into 40 × 40 square pixels and the dwell time of electron beam on each lattice is 0.001 s to minimize beam damage and cluster's drifting. All of the clusters in our experiment share the same measured parameters in STEM EELS except for the amplification factor.

**Details of the first principles calculations**

Density functional theory (DFT) and real-time time-dependent DFT (rt-TDDFT) were performed within general gradient approximations (GGA) in Perdew-Burke-Ernzerhof (PBE) implementation[70] using the PWmat code[68,71,72]. The norm-conserving pseudopotential produced by the code ONCVPSP[73] was used and $5d^{10}6s^1$ was considered to represent the valence electrons of the Au nanoclusters. The k-space was only sampled with the $\Gamma$ point. A plane-wave basis set with a cutoff of 45 Ry and the vacuum space of at least 15 Å were used in all calculations. The atomic coordinates of the Au nanocluster were optimized until the maximum force of all the atoms was less than 0.01 eV/Å.

For our rt-TDDFT, the N-electron system's time-dependent density is given by $n(r,t) = \sum_{j=1}^{N} |\psi_j(r,t)|^2$, where $\psi_j(r,t)$ is the single-particle occupied state. To solve the time-dependent Kohn-Sham (KS) single-particle equation $H_{KS}(n(r,t),t)\psi_j(r,t) = i\frac{\partial \psi_j(r,t)}{\partial t}$, the KS orbitals $\psi_j(r,t)$ are expanded by the adiabatic KS orbitals $\varphi_i(r,t)$: $\psi_j(r,t) = \sum_i C_{j,i}(t)\varphi_i(r,t)$. $C_{j,i}(t)$ is the expansion coefficient and $\varphi_i(r,t)$ satisfies $H_{KS}(n(r,t),t)\varphi_i(r,t) = \varepsilon_i(t)\varphi_i(r,t)$, where $\varepsilon_i(t)$ is the energy of $\varphi_i(r,t)$.

In our rt-TDDFT calculations, the ionic positions are fixed and $n(r,t)$ evolves with a time step of 0.01fs. A Dirac delta electric pulse polarized in the x direction is applied to obtain the absorption spectrum, in which the total simulation time length is 30fs. In order to distinguish the excitation modes, we applied a laser electric field shaped by a Gaussian wavepacket: $E(t) = E_x \sin(\omega t)\exp(\frac{-(t-t_0)^2}{\sigma^2})$, where $E_x$ is along the x direction and the intensity is $0.0257\ V/Å$, which can promise the optical response in the linear region, $t_0$ is 9 fs, $\sigma$ is 3.3 fs and $\omega$ is the resonant frequency. The total simulation time is more than 50 fs.

**The data that support the findings of this study are available from the corresponding author upon reasonable request.**

**Acknowledgments**

The authors gratefully acknowledge the financial support of the National Key R&D Program of China (2017YFA0303203, 2016YFA0203500), the National Natural Science Foundation of China (11904165, 11904166, 11934007, 11874407, 11774429, U1732273, U1830206 and U1732159), the Natural Science Foundation of Jiangsu Province (BK20160659 and BK2019028), the Strategic Priority Research Program of Chinese Academy of Science (Grant No. XDB 30000000), the Major Research plan of the National Natural Science Foundation of China (Grant No. 91961101) and the Fundamental Research Funds for the Central Universities.

The authors are grateful to the Pico Center at SUSTech, supported by the Presidential Fund and Development and Reform Commission of Shenzhen Municipality.

**Author contributions**

F. S. conceived and supervised the whole work. S. L., L, C. and K. H. took charge of sample preparation. L. X., S. L., K. H. and L. C. carried out the STEM EELS measurements and data analysis. K. L., J. H and J. D. took charge of the rt-TDDFT calculations. R. C. and J. C performed the classical model calculations. F. S., S. L. and K. L. wrote the manuscript. All the authors are invited to comment on the manuscript. and X. W., J. H., Q. D., X. W., Z. W. and G. W. participated in numerous discussions related to it.

S. L, L. X and K. L contribute equally to this work.

**Additional information**

Competing interests: The authors declare no competing financial interests.

**Figure Captions**

**Figure 1. Acquisition of high-quality plasmonic signals of individual clusters using STEM-EELS. a:** EELS mapping images of atomically precise gold clusters with $N$ varying from 70000 to 100. **b:** The electron energy loss (EEL) spectra acquired from the orange spots of an individual gold nanoparticle as shown in the inset. Gold's plasmon feature at ~2 - 3 eV can be probed. **c:** EEL spectra of $Au_{1100}$ cluster at different camera lengths (CL). At CL = 73 mm (the orange curve), we observe a plasmon resonance peak while there is no obvious signal at CL = 30 mm (the blue curve). **d:** The EEL spectra of individual gold atomically precise clusters collected from different positions as marked by spots with the same colors in the inset. For a big cluster ($Au_{70000}$), both BPP and SPP can be excited in the center, while only SPP can be excited at the edge. However, in a small atomic cluster ($Au_{600}$), the spectra obtained from the center and edge are similar.

**Figure 2. $N$ dependent evolution of the plasmon peaks. a:** the SPP evolution of the energy loss spectra collected at the edge of all the gold clusters from $Au_{70000}$ to $Au_{100}$, where the dashed line shows the variation of the center of the SPP peaks. **b:** BPP evolution of the energy loss spectra collected in the centers of the gold clusters from $Au_{70000}$ to $Au_{500}$, where the BPP remains steady when $N$ decreases from 70000 to 1100 and disappears abruptly for smaller clusters with $N < 887$. **c:** The exciting probability of BPP ($P_{BPP}$) plotted against $ln(N)$. It drops to nearly zero at $N = 887$. **d:** Scaling law of the SPP exciting probability ($P_{SPP}$), which is plotted in a $ln(P_{SPP}) \sim ln(N)$ double

logarithmic coordination. The slope of the linear fitting curve is 0.669, indicating $P_{SPP}$ ~ $N^{0.669}$. If $N$ ~ $R^3$ (R: radius), this roughly means an interesting rule $P_{SPP}$ ~ $R^2$. **e - g**: $N$ dependent evolution of the SPP position (**e**), ratio of the exciting probability of BPP and SPP ($P_{BPP} / P_{SPP}$) (**f**) and the FWHM of SPPs (**g**). Three characteristic regimes of gold cluster can be found from the data, including a slight red shift of SPP in regime 3; diminishing BPP and monotonic blue shift of SPP in regime 2, and multiple peaks of SPP in regime 1.

**Figure 3. From electron-boundary scattering modified classical plasmon to QCC plasmon below Au$_{887}$. a:** Schematic diagram of effective electron mean free path in an infinite bulk metal (left) and a finite cluster with diameter R (right). The large yellow circles represent the positive ion background while the small red ones represent the "free" electrons. In a finite cluster, the electron-boundary scattering becomes increasingly significant. **b:** Fitting the large-scale SPP red shift (experimental data: black, $N$ = 887-70000) with the classical plasmon theory after considering the boundary scattering. Two fitting lines marked in red and blue with A = 0.4 and 0.5 respectively show good agreement. **c:** Fitting the SPP blue shift ($N$ = 300 – 887) using the formula considering the quantum confinement effect.

**Figure 4. Superimposed transitions between quantized molecule-like electronic structures. a:** The Au$_{116}$ Dh geometric structures optimized by DFT. **b:** The calculated energy level diagram of Au$_{116}$ Dh structure. The orange solid line at 0

represents Fermi level. **c:** The dipole moment oscillations of $Au_{116}$ (red line, right part) as a function of time under the excitation of laser (black line, left part). **d, e:** Absorption spectra of $Au_{116}$ by rt-TDDFT calculation and experimental EEL spectrum of $Au_{100}$. Both theoretical and experimental results show three split features rather than a single SPP. **f - i:** To identify the excitation mode for different absorption peak (marked below) in $Au_{116}$ cluster, we check the corresponding time-dependent transition coefficient $|C_{j,i}(t)|^2$. The j and i denote the unoccupied and occupied energy level, respectively. The energy difference between j and i is in the range from $E_{abs}$-0.15 eV to $E_{abs}$+0.15 eV. Several important transitions are plotted. Only the transition with lowest energy of about 1.5 eV shows a plasmon-like behavior (f) while the other 3 (g - i) are all single electron excitation. The situation changes while we apply a larger field (see supplementary Figure. S2, 3).

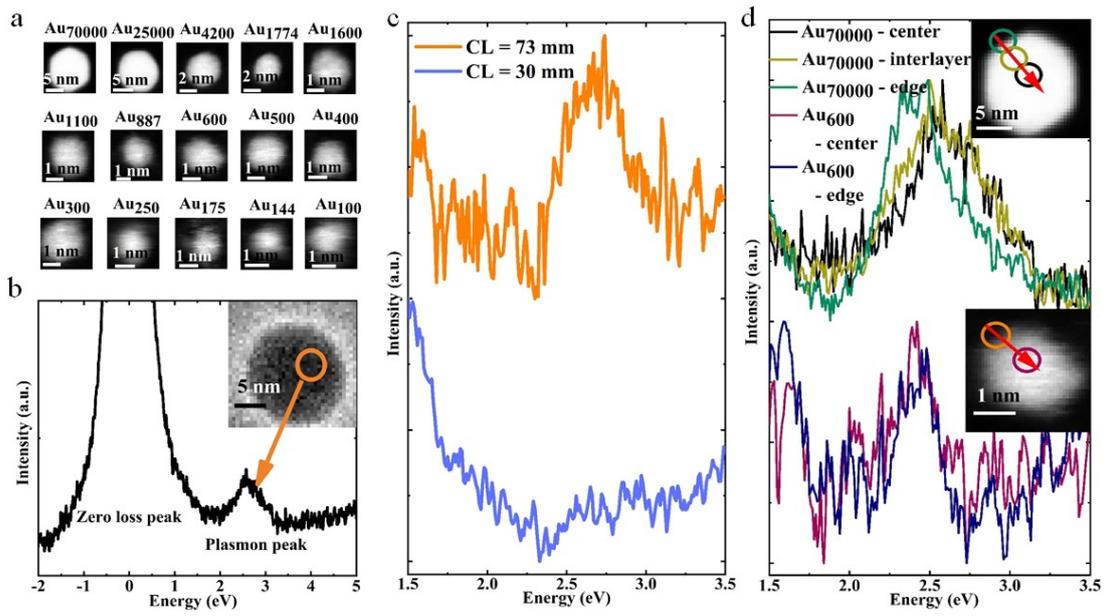

**Figure 1. Acquisition of high-quality plasmonic signals of individual clusters by STEM-EELS**

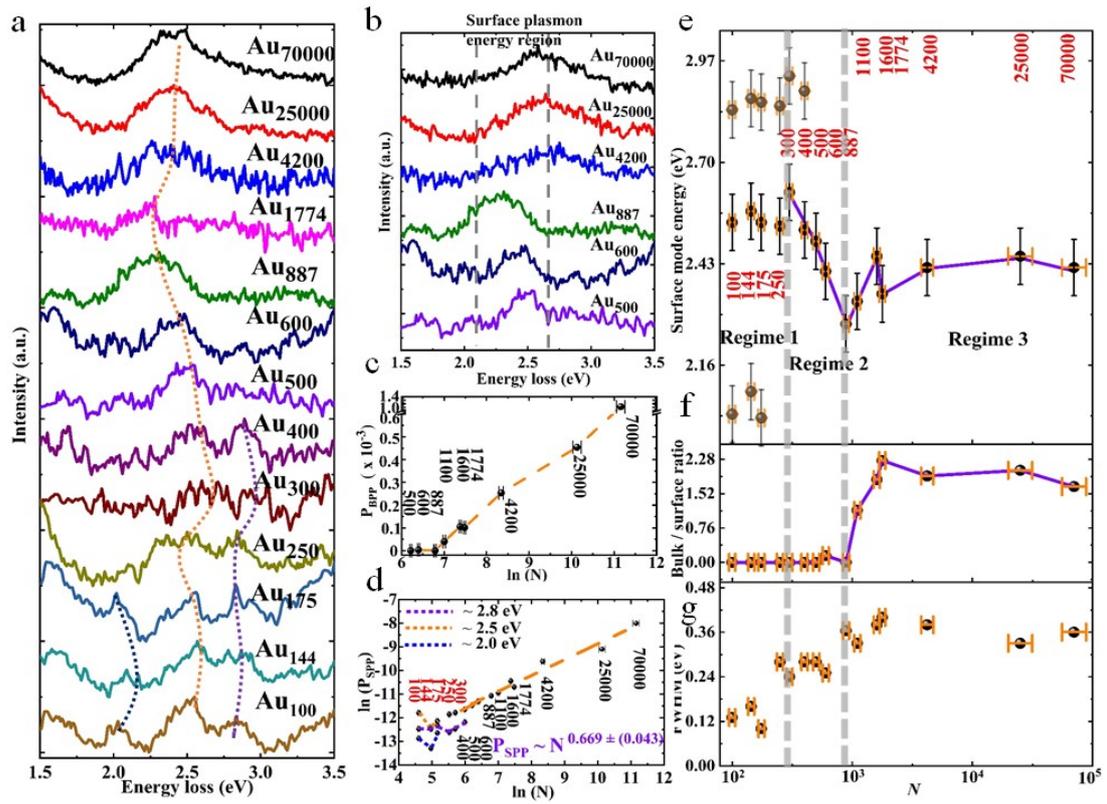

**Figure 2.** *N* dependent evolution of the plasmon peaks

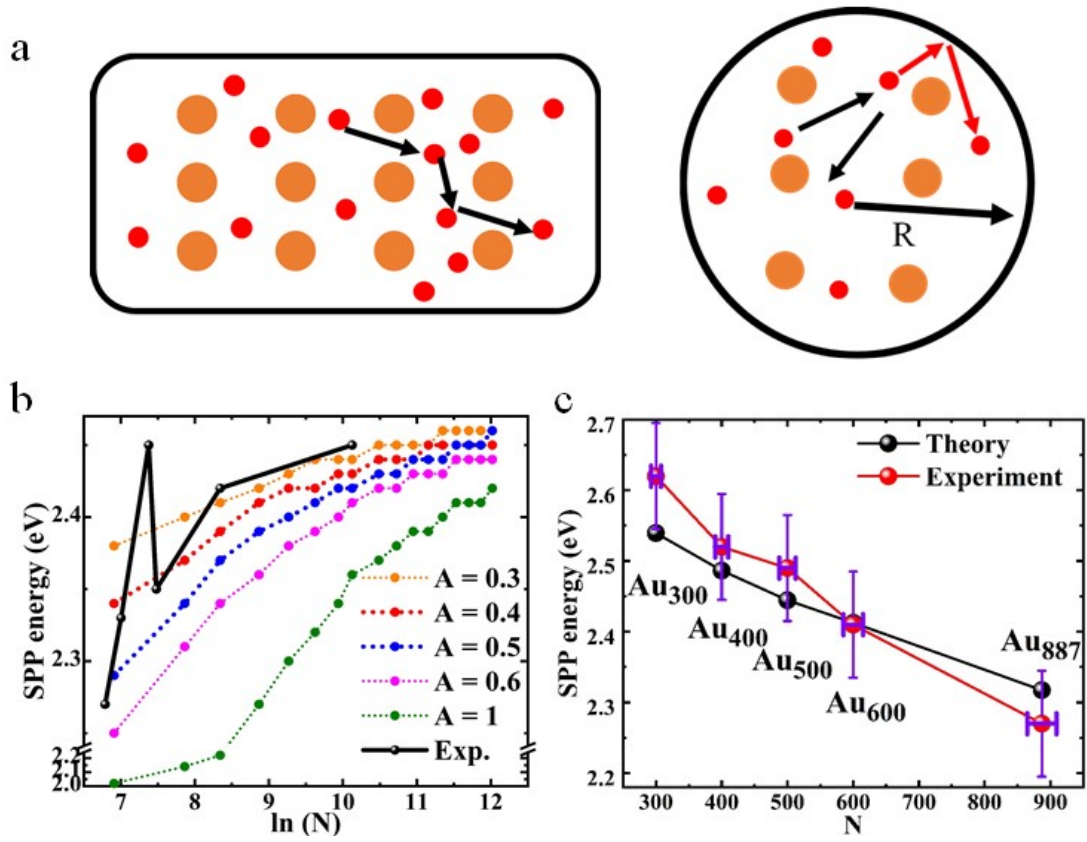

**Figure 3. From electron-boundary scattering modified classical plasmon to QCC plasmon below Au$_{887}$**

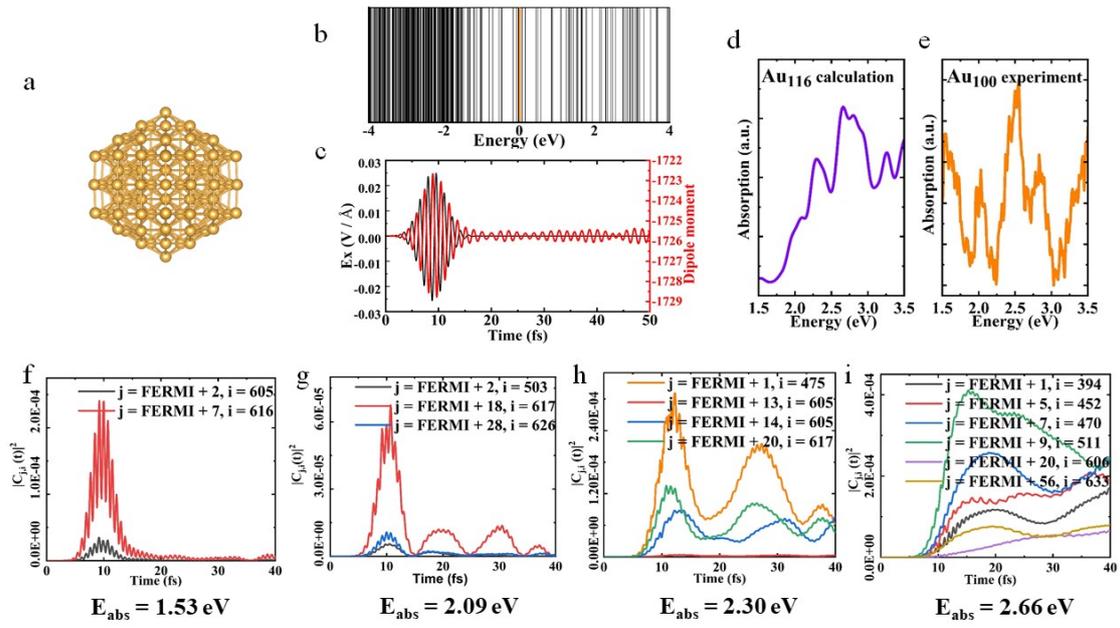

**Figure 4. Superimposed transitions between quantized molecule-like electronic structures**